\def\0uu{{\bf 0}}

\def\ruu{{\bf r}}

\def\Euu{{\bf E}}

\def\Ruu{{\bf R}}
\def\Huu{{\bf H}}

\def\Suu{{\bf S}}

\def\eps{{\epsilon}}

\def\alp{{\alpha}}

\def\pal{{\partial}}

\documentclass[dvipdfm,twocolumn,showpacs,preprintnumbers,amsmath,amssymb,prl]{revtex4}

\usepackage[dvipdfm]{graphicx}
\usepackage{dcolumn}
\usepackage{bm}

\begin{document}

\title{\bf Variational Analysis for Photonic Molecules} 
\author{Bin-Shei Lin }
\affiliation{Division of Nanoscience, National Center for
 High-Performance Computing, Hsinchu 300, Taiwan}
\vspace{8mm}
\begin{abstract}
\noindent

A new type of artificial molecule is proposed, which consists of
coupled defect atoms in photonic crystals, named as photonic
molecule. Within the major band gap, the photonic molecule
confines the resonant modes that are closely analogous to the
ground states of molecular orbitals. By employing the variational
theory, the constraint determining the resonant coupling is
formulated, that is consistent with the results of both the
scattering method and the group analysis. In addition, a new type
of photonic waveguide is proposed that manipulates the mechanism
of photon hopping between photonic molecules and offers a new
optical feature of twin waveguiding bandwidths.
\end{abstract}
 \maketitle

\vspace{4mm}

In the past decade, photonic defects have attracted much attention
owing to their scientific and technological applications in the
realization of high-Q microcavities or high transmittance
waveguides (WGs) \cite{ya0,oz,vi,me,me2,lin,bo}. A defect atom can
be embeded in a photonic crystal by perturbing the dielectricity
of a selected crystal ``atom'', that has photons with certain
frequencies locally trapped within the band gap of the surrounding
crystal structure. If defect atoms are embeded by design to form
the so called line defect then, within the band gap, it may
provide a mechanism of photon propagation via hopping from one
defect to its neighbors with a high transmittance
\cite{ya,ba1,ba2}. Consequently, the integrated optical circuits
of functional elements can be realized through skillful
integration of photonic defects and WGs, and is expected to offer
potential applications in telecommunication \cite{ko, mcg}.

A point of importance but of much less attention is that the
formation of photonic WGs are conventionally considered as
arranging the desired defect atoms along a line, however, this
approach has limited the potential of development. We suggest a
broader vision of the manipulation of photonic defects through the
investigation of photonic molecules that are defined as follows.
In photonic crystals, the defect atoms are closely arranged to
form a structure that is similar to a real molecule $-$ So named
because, within the major band gap, the resonant modes confined by
such a structure are closely analogous to the ground states of the
molecular orbitals (MOs) of the real molecule. For example, Fig. 1
(a) shows a photonic molecule named as the photonic benzene, whose
defect modes are shown in Fig. 3. By employing the variational
theory, the constraint determining the resonant coupling of
photonic molecule is formulated, that is consistent with the
results from those of both the scattering solution and the group
analysis. In particular, manipulating the mechanism of photon
hopping between photonic benzenes can provide the function of
guiding photons along the benzene chain with a high transmittance,
and presents a new optical feature of twin waveguiding bandwidths,
as shown as Fig 4.

Because of the importance on interpreting the mechanism of defect
coupling, there are mainly two solid-state theoretical approaches,
the tight-binding (TB) \cite{li,ya,ba1} and Wannier function
methods \cite{leu,al,ga}, have been applied to study the coupled
cavities. The photonic version of TB method extends the idea of
linear combination of atomic orbitals (LCAO), in which the defect
modes are analogous to the atomic wave function, and suppose that
only the nearest-neighbor couplings are relevant to find the
dispersion relation for waveguide mode. For the latter, the
localized defect modes are expanded by Wannier functions to
calculate their intensity variations, where the Wannier functions
are calculated by plan-wave method or TB method coupled with
supercell approximation. Here, another powerful approach of
variational analysis is introduced for many defect-atoms system.

Considering first an electric resonant mode $\Euu_d(\ruu)$ of a
single defect in a finite-size photonic crystal, the Maxwell
equations obeyed by $\Euu_d(\ruu)$ can be further simplified as
\begin{equation}
\hat{\Huu}\Euu_d(\ruu)=\eps_d(\ruu)(\omega_d/c)^2\Euu_d(\ruu),\label{e1}
\end{equation}
where the operator $\hat{\Huu}$ is defined as $-\nabla^2$ for the
2D system or $\nabla\times\nabla\times$ for the 3D system. Also,
$\eps_d(\ruu)$ denotes the dielectric constant of the single
defect system, and $\omega_d$ is the eigenfrequency of the
eigenmode $\Euu_d(\ruu)$. Those modes occure within the major band
gap are most concerned in this paper, and can be taken as real and
normalized by
\begin{figure}[b]
\leavevmode \center{
\includegraphics[width=8.8cm,height=3.6cm]{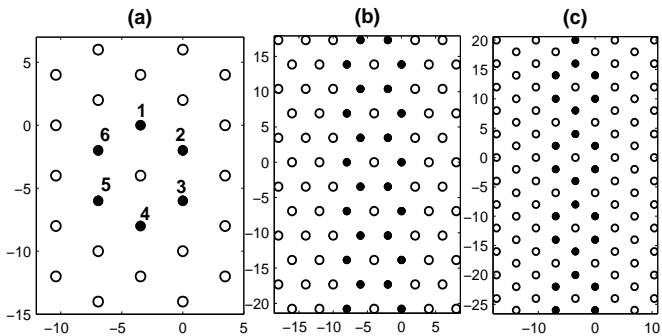}
\caption{Formation of (a) photonic benzene, (b) benzene WG of
$\pi-$type and (c) benzene WG of $\sigma-$type in a 2D hexagonal
crystal; each solid dot denotes a defect atom
 of radius $\rho$.}
 }
\end{figure}
$$<\Euu_d(\ruu)|\eps_d(\ruu)|\Euu_d(\ruu)>\stackrel{\mathrm{def}}{=}
\int_{domain}\eps_d(\ruu)\Euu_d(\ruu)\cdot\Euu_d(\ruu)d\ruu=1.$$

For the same photonic crystal considered in (\ref{e1}) but
embedded with a photonic molecule, the allowed resonant modes are
assumed as a superposition of the individual defect mode.
Basically, it is analogous to the idea of LCAO, namely

\begin{equation}
\Euu_n(\ruu)=\sum_{i=1}^{n_d}C_{ni}\Euu_d(\ruu-\Ruu_i),\label{e2}
\end{equation}
where $\Euu_n(\ruu)$ is the nth resonant mode of the photonic
molecule, $n_d$ the number of defect atoms, $C_{ni}$ the
undetermined coefficient, and $\Ruu_i$ the coordinates of the ith
defect atom. Similarly, $\Euu_n(\ruu)$ satisfies (\ref{e1}) but
with $\eps_d(\ruu)$ replaced by the dielectric constant
$\eps_{pm}(\ruu)$ of the photonic molecule system, and $\omega_d$
replaced by the frequency $\omega_n$ of the eigenmode
$\Euu_n(\ruu)$. That is

\begin{equation}
\hat{\Huu}\Euu_n(\ruu)=\eps_{pm}(\ruu)(\omega_n/c)^2\Euu_n(\ruu).\label{e3}
\end{equation}

Equation (\ref{e2}) associated with (\ref{e3}) is a linear
variational problem. Assigning different coefficient $C_{ni}$ to
each mode $E_d$ may create different $E_n$, but the structure of
photonic molecule will decide which resonant modes are allowed.
This inference will be reflected on restricting $C_{ni}$ to
satisfy the minimum of functional frequency, defined as

\begin{equation}
\Omega(C_{ni})\stackrel{\mathrm{def}}{=}\frac{\left<\Euu_n(\ruu)\left|\hat{\Huu}\right|\Euu_n(\ruu)\right>}
{\left<\Euu_n(\ruu)|\eps_{pm}(\ruu)|\Euu_n(\ruu)\right>}
=\frac{\sum_{ij}C_{ni}C_{nj}\Huu_{ij}}{\sum_{ij}C_{ni}C_{nj}\Suu_{ij}}.
\nonumber
\end{equation}
Namely, $\Omega$ is equivalent to the familiar Rayleigh-Ritz
principle. Here $\Huu_{ij}$ and $\Suu_{ij}$ denote, respectively,
the elements of the Hamiltonian matrix and the overlap matrix.
According to (\ref{e1}) and (\ref{e2}), $\Huu_{ij}$ can be written
as
\begin{eqnarray}
&&\Huu_{ij}=\left<\Euu_d\left(\ruu-\Ruu_i\right)\left|\hat{\Huu}\right|\Euu_d\left(\ruu-\Ruu_j\right)\right>\nonumber\\
&=&\begin{cases}\left(\omega_d/c\right)^2=\alp, \ \  \mbox{for
$i=j$};\cr \left(\omega_d/c\right)^2\beta_1, \ \mbox{for $(i,j)$
being the nearest-neighbor}; \cr \left(\omega_d/c\right)^2\beta_2,
\ \mbox{for $(i,j)$ being the second-neighbor}; \cr
\left(\omega_d/c\right)^2\beta_3, \ \mbox{for $(i,j)$ being the
third-neighbor} ,
\end{cases}
\nonumber
\end{eqnarray}
where
$\beta_{(ij)}=<\Euu_d(\ruu-\Ruu_i)|\eps_d(\ruu-\Ruu_j)|\Euu_d(\ruu-\Ruu_j)>$
denotes the hopping integral, whose magnitude measures the
coupling strength and decays rapidly with increasing the distance
$|\Ruu_i-\Ruu_j|$, i.e. $|\beta_1|>|\beta_2|>|\beta_3|$ (the more
the defect sites split, the weaker their coupling \cite{ta}).
Therefore, hopping terms can be classified according to the
separation of the coupled defects. Here, only three relevant
hopping terms are considered. Moreover, under the assumption that
each individual defect mode is highly localized around its defect
site, the field overlap between different defect atoms is small
and the overlap integral $\Suu_{ij}$ can thus be approximated as

\begin{equation}
\Suu_{ij}=<\Euu_d(\ruu-\Ruu_i)|\eps_{pm}(\ruu)|\Euu_d(\ruu-\Ruu_j)>\approx\delta_{ij}.
\label{app}
\end{equation}

Now, vary $C_{ni}$ to minimize the functional frequency $\Omega$,
with the necessary condition of $\pal \Omega/\pal C_{ni}=0$,
$i=1,2,\dots,n_d$. One can obtain
\begin{equation}
\sum_{j=1}^{n_d}\left[\Huu_{ij}-\left(\omega_n/c\right)^2\delta_{ij}\right]C_{nj}=0.\nonumber
\end{equation}
The constrain of resonant frequencies can thus be derived from the
solvability condition of $C_{ni}$. This leads to

\begin{equation}
\det\left[\Huu_{ij}-\left(\omega_n/c\right)^2\delta_{ij}\right]=0.
\label{e6}
\end{equation}

Equation (\ref{e6}) indicates that the allowed resonant
frequencies in a given photonic molecule are dominated by hopping
integrals. Furthermore, these hopping terms are dependent upon the
dielectric structure of the photonic molecule. Therefore, every
resonant mode is characterized by photonic molecule and exhibits
different optical transmittance. To check the accuracy of Eq.
(\ref{e6}), we first apply Eq. (\ref{e6}) to the photonic benzene,
that yields
\begin{eqnarray}
\left| \begin{matrix} \alp-\gamma & \tilde{\beta}_1 & \tilde{\beta}_2 & \tilde{\beta}_3 &
\tilde{\beta}_2 & \tilde{\beta}_1 \cr
\tilde{\beta}_1&\alp-\gamma&\tilde{\beta}_1&\tilde{\beta}_2&\tilde{\beta}_3&\tilde{\beta}_2 \cr
\tilde{\beta}_2&\tilde{\beta}_1&\alp-\gamma&\tilde{\beta}_1&\tilde{\beta}_2&\tilde{\beta}_3 \cr
\tilde{\beta}_3&\tilde{\beta}_2&\tilde{\beta}_1&\alp-\gamma&\tilde{\beta}_1&\tilde{\beta}_2 \cr
\tilde{\beta}_2&\tilde{\beta}_3&\tilde{\beta}_2&\tilde{\beta}_1&\alp-\gamma&\tilde{\beta}_1 \cr
\tilde{\beta}_1&\tilde{\beta}_2&\tilde{\beta}_3&\tilde{\beta}_2&\tilde{\beta}_1&\alp-\gamma
\end{matrix} \right|=0,
\label{e7}
\end{eqnarray}
where we let $\gamma=(\omega_n/c)^2$ and
$\tilde{\beta}_i=(\omega_d/c)^2\beta_i$ for simplification. The
determinant in Eq. (\ref{e7}) is called the 6th-order circulant,
and is equivalent to

\begin{equation}
\prod_{n=1}^6\left[\left(\alp-\gamma\right)+e_n\tilde{\beta}_1+e_n^2\tilde{\beta}_2+e_n^3
\tilde{\beta}_3+e_n^4\tilde{\beta}_2+e_n^5\tilde{\beta}_1\right]=0,
\nonumber
\end{equation}
where $e_1,e_2\dots e_6$ are the six roots of unity, i.e.
$e_n=exp(2\pi i n/6)$. Hence, for $n=1\dots 6$, the frequencies
$\omega_n$ of the six resonant modes $\Euu_n$ can be found as
\begin{equation}
\omega_n=\omega_d\sqrt{1+2\cos\frac{2\pi n}{6}\beta_1+2\cos\frac{2\pi n}{3}\beta_2+(-1)^n\beta_3}.
\label{e11}
\end{equation}

Obviously, Eq. (\ref{e11}) indicates that there are two
doubly-degenerate frequencies with $n=1,5$ and $n=2,4$, and two
nondegenerate frequencies with $n=3$ and $n=6$, thus, four high-Q
resonant frequencies will occur within the major band gap. From
the view point of the symmetry group , the photonic benzene
belongs to the point group $C_{6v}$ for the 2D or $D_{6h}$ for the
3D systems, whose irreducible representation $\Gamma$ on defect
sites can be reduced to the decomposition $\Gamma=A_1+B_2+E_1+E_2$
or $\Gamma=A_{2u}+B_{2g}+E_{1g}+E_{2u}$, respectively. Exactly
speaking, it again illustrates two doubly degenerate modes of
$E_{1}$ (or $E_{1g}$) and $E_{2}$ (or $E_{2u}$), and two
nondegenerate modes of $A_{1}$ (or $A_{2u}$) and $B_{2}$ (or
$B_{2g}$). Furthermore, these predictions are also consistent with
the numerical solution of Eq. (\ref{e3}), that is calculated by
scattering method (cf. \cite{ta}). The resultant transmission for
the 2D case is plotted in Fig. 2, and the allowed resonant modes
are shown in Fig. 3, where we consider a 2D finite-size hexagonal
photonic crystal with a dielectric contrast ratio $8.4/1.0$
(rod/background) and a radius-to-spacing ratio $0.4/4.0$. The
defect radius $\rho$ is taken as zero.

Figure 2 shows four nomalized resonant frequencies $0.369$,
$0.375$, $0.455$ and $0.508$, that split from $\omega_d=0.419$ due
to the defect coupling in a photonic benzene. By substituting
these five frequencies into Eq. (\ref{e11}), the hopping terms
$\beta_1$, $\beta_2$ and $\beta_3$ can be calculated by
least-squares method, and they are $0.178$, $0.051$ and $-0.010$.
Table I summarizes four cases of defect sizes, in which the all
values of hopping parameters achieve the accuracy of two decimal
places for Eq. (\ref{e11}). It clearly shows that the larger the
defect radius is, the smaller hopping parameters. This means that
the defect couplings in photonic benzene become weaker as the
defect radius is increased. Moreover, when the defect radius
$\rho$ is increased up to about $0.2$, the property of four
transmission peaks disappeares, owing to that the shallow
perturbation of dielectricity for defect atoms will create shallow
modes \cite{ya0}.

\begin{table}[b]
\caption{The calculated values of normalized resonant frequencies
and hopping parameters for photonic benzene with different
$\rho$.}
\begin{tabular}{ccccccccc}\hline\hline
$\rho$ & $\omega_d$ & $\omega_1$ &  $\omega_2$ & $\omega_3$ &
$\omega_4$ & $\beta_1$ & $\beta_2$ & $\beta_3$ \\ \hline
0.0&0.419&0.455&0.375&0.369&0.508&0.178&0.051&-0.010\\
0.05&0.417&0.452&0.373&0.367&0.505&0.177&0.051&-0.009\\
0.1&0.409&0.442&0.369&0.361&0.493&0.171&0.048&-0.006\\
0.15&0.395&0.423&0.361&0.351&0.468&0.155&0.040&-0.002\\ \hline
\hline
\end{tabular}
\end{table}
In fact, Eq. (\ref{e3}) is equivalent to the effective
Shr$\ddot{o}$dinger equation of H$\ddot{u}$ckel $\pi-$electron
theory (developed in 1931 \cite{hu}), if we make the resonant
modes of 3D photonic benzene be analogous to the $\pi-$electrons
of benzene molecule. However, Eq. (\ref{e3}) is much simpler.
Theoretically, the $\pi$-electrons arise from a planar unsaturated
organic molecule whose MOs can be divided into the $\sigma$ and
$\pi$ MOs according to the reflection symmetry in the molecular
plane. Both systems belong to the same point group $D_{6h}$ and
have the same degeneracy (cf. \cite{at}, p.261), but possess
completely different meanings. Conceptually, the photonic molecule
acts as a perfect model of artificial molecule, since the resonant
modes are much easier to be realized than the bonding orbitals of
real molecule. Similar phenomena can also be found in the
quantum-dot molecules \cite{pi} or the coupled pairs of GaAs
cavities (note that these systems are also termed as photonic
molecule, cf. Bayer et al. \cite{bayer}). Fig. 3 shows the lowest
resonant modes allowed in a 2D photonic benzene with $\rho=0.0$,
and they are labelled according to the symmetry species of the
group $C_{6v}$, which are analogous to the six $\pi$ MOs of
benzene molecule but is lacking of the $C_{1h}$ symmetry.
\begin{figure}[t] \leavevmode \center{
\includegraphics[width=9.5cm,height=4.5cm]{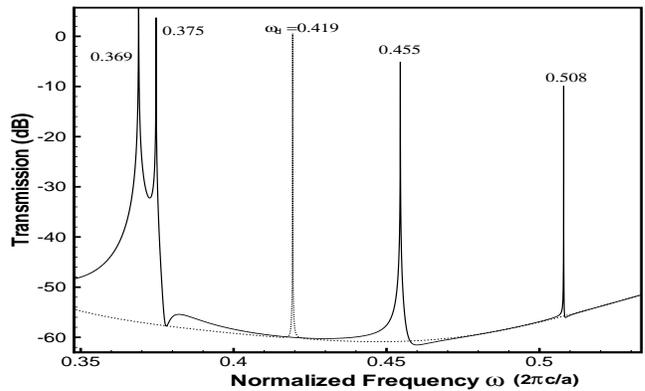}
\caption{Transmission amplitude of electric field as function of
normalized frequency for a photonic benzene (solid curve) and a
defect atom (dotted curve). The marked values are four frequencies
of the high-Q modes which are split from the $\omega_d$ due to
defect coupling and fallen in the major band gap whose rang is as
shown in Fig 4.}}\label{f2}
\end{figure}
\begin{figure}[b]
\leavevmode \center{
\includegraphics[width=9.3cm,height=7.5cm]{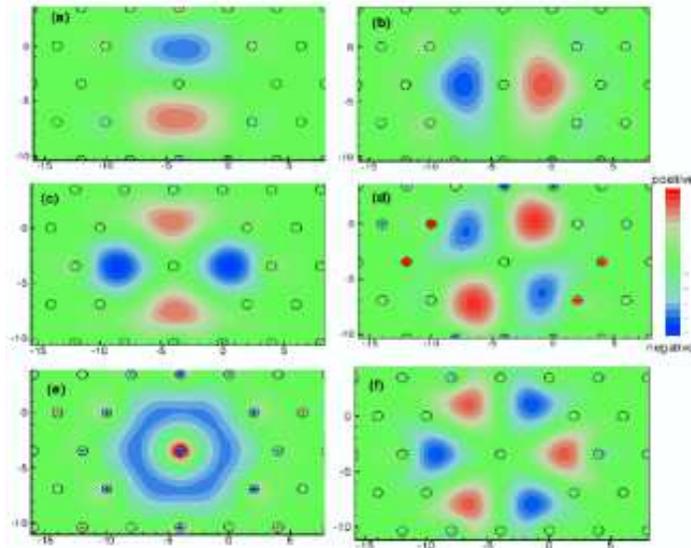}
\caption{Resonant electric field pattern in a photonic benzene of
$\rho=0$ for the six lowest resonant modes with normalized
frequencies of (a) 0.369 with $E_{1}$ mode, (b) 0.369 with $E_{1}$
mode, (c) 0.455 with $E_{2}$ mode, (d) 0.455 with $E_{2}$ mode,
(e) 0.375 with $A_{1}$ mode and (f) 0.508 with $B_{2}$ mode,
respectively. In general, the more node, the higher frequency.}
 }
\end{figure}
\begin{figure}[htp]
\leavevmode \center{
\includegraphics[width=8.9cm,height=7.2cm]{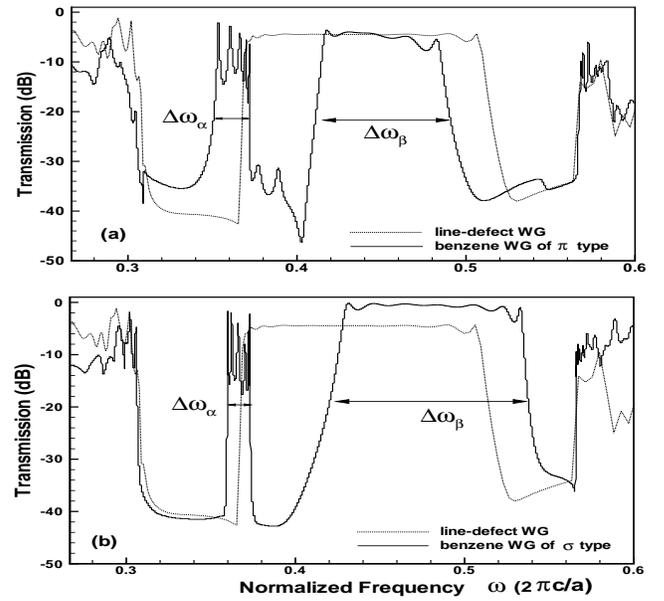}
\caption{Transmission created by benzene WGs of (a) $\pi-$type and
(b) $\sigma-$type with $\rho=0.0$. Both types present the feature of
twin waveguiding bandwiths, marked as $\triangle\omega_{\alp}$ and
$\triangle\omega_{\beta}$.}
 }
 \end{figure}
\begin{figure}[htp]
\leavevmode \center{
\includegraphics[width=3.6cm,height=8.5cm,,angle=-90]{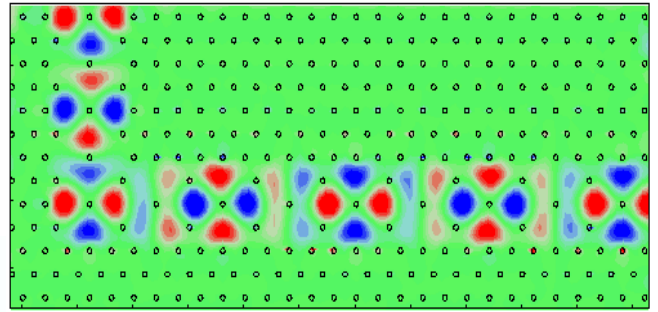}
\caption{Eletric field distribution of a TM light with $E_{2}$
mode travels through a $90^o$ bend in a benzene WG where the
junction connects a $\pi$ type WG with a $\sigma$ type; incident
wave with normalized frequencies 0.449 enters from the upper
left.}}\label{f5}
\end{figure}

Most importantly, by applying the modular concept of photonic
benzene can create photonic WGs, and we called these WGs as
benzene WGs. In the chemistry terminology, benzene WGs can be
classified as $\pi$ and $\sigma$ types corresponding to the
bonding types between two real benzene rings, as illustrated in
Fig. 1 (b) \& (c). It is remarkable that transmission of the
benzene WGs reveals the special feature of a twin waveguiding
bandwidths marked as $\triangle\omega_{\alp}$ and
$\triangle\omega_{\beta}$, where
$\triangle\omega_{\alp}<\triangle\omega_{\beta}$, as shown in Fig.
4. This means that the benzene WGs are able to provide two working
bandwidths at the same time. In addition, Fig. 5 shows that a TM
light with the $E_{2}$ mode travels through a $90^o$ bend from the
$\pi$ type to the $\sigma$ type. Of course, the same phenomenon
can also be observed in other modes.

In conclusion, we suggest a new and practicable idea for the
manipulation of photonic defects, that includes the so-called
photonic molecule and benzene WG. The optical properties of
photonic molecules has been investigated by variational theory,
which shows that the allowed resonant frequencies inside a
photonic molecule are dominated by hopping parameters through the
constraint (\ref{e6}). Taking the photonic benzene as an example,
six resonant modes with two doubly-degenerate and two
nondegenerate are found and verified by both of the scattering
method and group theory. Especially, the benzene WGs created by
the modular manipulation of photonic benzenes are demonstrated to
possess the interesting feature of a twin waveguiding bandwidths.
Namely, benzene WGs provide two working bandwidths at the same
time and make the function of guiding photons more flexible.

This work is supported in part by the {\em Global Fiberoptics,
Inc.}



\begin{thebibliography}{99}
\bibitem{ya0}
E. Yablonovitch, T. J. Gmitter, R. D. Meade, A. M. Rappe, K. D.
Brommer, and J. D. Joannopoulos, \prl {\bf 67}, 3380 (1991).
\bibitem{oz}
E. $\ddot{O}zbay$, G. Tuttle, M. Sigalas, C. M. Soukoulis, and K.
M. Ho, \prb {\bf 51}, 13961 (1995).
\bibitem{vi}
P. R. Villeneuve, S. Fan, and J. D. Joannopoulos, \prb {\bf 54}, 7837 (1996).
\bibitem{me}
A. Mekis, J. C. Chen, I. Kurland, S. Fan, P. R. Villenuve, and J. D. Joannopoulos,
\prl {\bf 77}, 3787 (1996).
\bibitem{me2}
A. Mekis, S. Fan, and J. D. Joannopoulos, \prb {\bf 58}, 4809 (1998).
\bibitem{lin}
S. Y. Lin, E. Chow, V. Hietala, P. R. Villenuve, and J. D.
Joannopoulos, Science {\bf 282}, 274 (1998).
\bibitem{bo}
S. Boscolo and M. Midrio, Opt. Lett. {\bf 27}, 1001 (2002).
\bibitem{ya}
A. Yariv, Y. Xu, R. K. Lee, and A. Scherer, Opt. Lett. {\bf 24}, 711 (1999).
\bibitem{ba1}
M. Bayindir, B. Temelkuran, and E. Ozbay, \prl {\bf 84}, 2140 (2000).
\bibitem{ba2}
M. Bayindir, B. Temelkuran, and E. Ozbay, \prb {\bf 61}, R11855 (2000).
\bibitem{ko}
H. Kosaka, T. Kawashima, A. Tomita, M. Notomi, T. Tamamura, T. Sato, and S. Kawakami,
Appl. Phys. Lett. {\bf 74}, 1370 (1999).
\bibitem{mcg}
A. R. McGurn, \pre {\bf 65}, 075406 (2002).
\bibitem{li}
E. Lidorikis, M. M. Sigalas, E. N. Economou, and C. M. Soukoulis, \prl {\bf 81}, 1405 (1998).
\bibitem{leu}
K. M. Leung, J. Opt. Soc. Am. B {\bf 10}, 303 (1993).
\bibitem{al}
J. P. Albert, C. Jouanin, D. Cassagne, and D. Bertho, \prb {\bf 61}, 4381 (2000).
\bibitem{ga}
A. Garcia-Martin, D. Hermann, K. Busch, and P. W$\ddot{o}$lfle,
Mat. Res. Soc. Symp. Proc. {\bf 722}, L 1.1.1 (2002).
\bibitem{ta}
G. Tayeb and D. Maystre, J. Opt. Soc. Am. A {\bf 14}, 3323 (1997).
\bibitem{hu}
E. H$\ddot{u}$ckel, Z. physik {\bf 70}, 204 (1931).
\bibitem{at}
P. W. Atkins and R. S. Friedman, {\sl Molecular Quantum Mechanics
} (Oxford University, Oxford, 1997).
\bibitem{pi}
M. Pi, A. Emperador, M. Barranco, F. Garcias, K. Muraki, S.
Tarucha, and D. G. Austing, \prl {\bf 87}, 066801 (2001).
\bibitem{bayer}
M. Bayer, T. Gutbrod, J. P. Reithmaier, A. Forchel, T. L.
Reinecke, P. A. Knipp, A. A. Dremin and V. D. Kulakovskii, \prl
{\bf 81}, 2582 (1998).
\end{thebibliography}
\end{document}